\documentclass[aps,prb,preprint,superscriptaddress]{revtex4-1}
\usepackage{graphicx}
\usepackage{amssymb}
\usepackage{amsmath}
\usepackage{color}

\begin{document}

\title{Fluctuating magnetism of Co- and Cu-doped NaFeAs}

\author{Jonathan Pelliciari}
\email[Author to whom correspondence should be addressed: ]{pelliciari@bnl.gov}
\affiliation{Photon Science Division, Paul Scherrer Institut, CH-5232 Villigen PSI, Switzerland}
\affiliation{NSLS-II, Brookhaven National Laboratory, Upton, NY 11973, USA}
\author{Kenji Ishii}
\affiliation{Synchrotron Radiation Research Center, National Institutes for Quantum and Radiological Science and Technology, Sayo, Hyogo 679-5148, Japan}   
\author{Lingyi Xing}
\affiliation{Beijing National Lab for Condensed Matter Physics, Institute of Physics, Chinese Academy of Sciences, Beijing 100190, China}
\author{Xiancheng Wang}
\affiliation{Beijing National Lab for Condensed Matter Physics, Institute of Physics, Chinese Academy of Sciences, Beijing 100190, China}
\author{Changqing Jin}
\affiliation{Beijing National Lab for Condensed Matter Physics, Institute of Physics, Chinese Academy of Sciences, Beijing 100190, China}
\affiliation{Collaborative Innovation Center for Quantum Matters, Beijing 100190, China}
\author{Thorsten Schmitt}
\email[]{thorsten.schmitt@psi.ch}
\affiliation{Photon Science Division, Paul Scherrer Institut, CH-5232 Villigen PSI, Switzerland}

\date{\today}

\begin{abstract}
We report an x-ray emission spectroscopy (XES) study of the local fluctuating magnetic moment ($\mu_{bare}$) in $\mathrm{NaFe_{1-x}Co_{x}As}$ and $\mathrm{NaFe_{1-x}Cu_{x}As}$.
In NaFeAs, the reduced height of the As ions induces a local magnetic moment higher than $\mathrm{Ba_2As_2}$, despite lower T$_N$ and ordered magnetic moment. As NaFeAs is doped with Co $\mu_{bare}$ is slightly reduced, whereas Cu doping leaves it unaffected, indicating a different doping mechanism: based on electron counting for Co whereas impurity scattering dominates in the case of Cu. 
Finally, we observe an increase of $\mu_{bare}$ with temperature in all samples as observed in electron- and hole-doped $\mathrm{BaFe_2As_2}$. Since both Co and Cu doping display superconductivity, our findings demonstrate that the formation of Cooper pairs is not connected with the complete loss of fluctuating paramagnetic moments.
\end{abstract}

\maketitle
\begin{figure}
\includegraphics[scale=1]{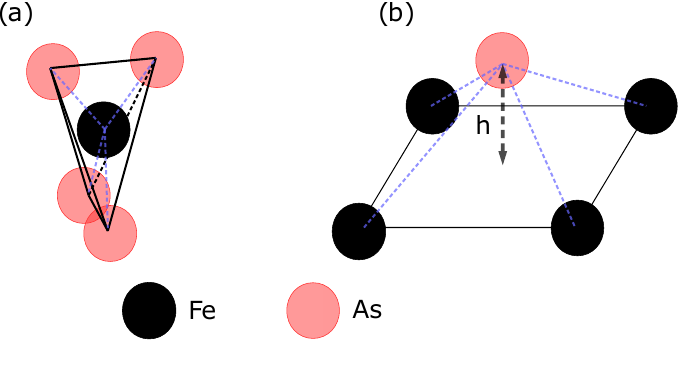}
\caption{\label{fig:fig1} (a) and (b) Building blocks of the FeAs layer. (a) Tetrahedral coordination of the Fe atoms. (b) Indication of the height \textit{h} of the As atoms from the Fe layer, which is a critical parameter for the magnetism of Fe pnictides.}
\end{figure}

The structure of superconducting Fe pnictides is composed of FeAs layers separated by spacing ions where Fe occupies a four-fold coordination site with a tetrahedral environment of As ions as depicted in Fig.~\ref{fig:fig1}(a). This coordination can alternatively be seen as an Fe checkerboard layer with As ions at the center of every single Fe square alternating above and below it, as illustrated in Fig.~\ref{fig:fig1}(b). A key parameter for the magnetism of Fe pnictides is the height (\textit{h}) of the As ions with respect to the Fe layer (Fig.~\ref{fig:fig1}(b)) \cite{zhang_effect_2014,zhang_electron_2016,carr_electron_2016,pelliciari_presence_2016,pelliciari_intralayer_2016,johnson_iron-based_2015,dai_antiferromagnetic_2015}.
In NaFeAs, the large \textit{h} (1.416~\AA) induces magnetic frustration compared to BaFe$_2$As$_2$ (1.358~\AA) \cite{zhang_effect_2014}, leading to the reduction of the ordered magnetic moment and T$_{N}$ (see Table \ref{tab:table1} for the respective values). \cite{li_structural_2009,johnston_puzzle_2010,stewart_superconductivity_2011,zhang_effect_2014,pelliciari_presence_2016,pelliciari_intralayer_2016}. 
Despite the reduced ordered moment, the spin excitations of NaFeAs have been detected by both inelastic neutron scattering (INS) and resonant inelastic x-ray scattering (RIXS) \cite{zhang_effect_2014,pelliciari_intralayer_2016,carr_electron_2016}, and from the integration in energy and momentum spaces of the INS signal a fluctuating magnetic moment higher than $\mathrm{BaFe_2As_2}$ has been quantified \cite{zhang_effect_2014}. These evidences are a clear demonstration of the importance of magnetic fluctuations in the NaFeAs  (111) series.

\begin{table}
\caption{\label{tab:table1} Summary of T$_N$, $\mu_{ord}$, and \textit{h} for BaFe$_2$As$_2$ and NaFeAs. Reproduced with permission from S. Li \textit{et al.}, "Structural and magnetic phase
transitions in Na$_{1-\delta}$FeAs", Physical Review B \textbf{80}, 020504 (2010); and G. R. Stewart, "Superconductivity in iron compounds", Reviews of Modern Physics \textbf{83}, 1589 (2011). Copyright 2010-2011 by the American Physical Society. Reproduced with permission from D. C. Johnston Advances in Physics \textbf{59}, 803 (www.tandfonline.com). \cite{li_structural_2009,johnston_puzzle_2010,stewart_superconductivity_2011}}
\begin{ruledtabular}
\begin{tabular}{lcr}
 & BaFe$_2$As$_2$ & NaFeAs\\
\hline
T$_N$ & 140~K & 45~K\\
$\mu_{ord}$ & 1.3 $\mu_B$ & 0.1 $\mu_B$\\
\textit{h} & 1.358 \AA & 1.416 \AA\\
\end{tabular}
\end{ruledtabular}
\end{table}

A peculiarity of Fe pnictides, in respect to the cuprates, is the high flexibility to achieve SC through different types of doping. Fe pnictides can be electron-, hole-, or isovalent-doped with SC emerging in all the cases \cite{johnson_iron-based_2015,johnston_puzzle_2010,stewart_superconductivity_2011,hosono_iron-based_2015,shibauchi_quantum_2014}. The doping flexibility is not only limited to the type of carriers (electrons or holes) but also to the site of the dopant atoms. Dopants can be placed in all the structural sites: the spacing layer (\textit{e.g.} $\mathrm{Ba_{1-x}K_xFe_2As_2}$), the Fe layer (\textit{e.g.} $\mathrm{NaFe_{1-x}Co_xAs}$), and the As layer [\textit{e.g.} $\mathrm{BaFe_2(As_{1-x}P_x)_2}$] \cite{johnson_iron-based_2015,johnston_puzzle_2010,stewart_superconductivity_2011,hosono_iron-based_2015,shibauchi_quantum_2014}. The effect of doping transition metals such as Co, Ni, and Cu into the Fe layer has been studied with several techniques sensitive to the electronic structure such as angle resolved photo-emission experiments (ARPES) \cite{van_roekeghem_spectral_2016,ye_extraordinary_2014,richard_fe-based_2011,richard_angle-resolved_2009,ideta_dependence_2013,ge_anisotropic_2013,neupane_electron-hole_2011,lu_arpes_2009}. However, a general agreement on the doping effect on the electronic structure has not been reached since in some cases the Fermi level is rigidly shifted \cite{cui_arpes_2013,richard_fe-based_2011,van_roekeghem_spectral_2016}, whereas in other cases doping involves an enhancement of impurity scattering additionally to a rigid shift of the Fermi level \cite{wadati_where_2010,kim_effects_2012,ideta_dependence_2013}.

\begin{figure*}
\includegraphics[ scale=1]{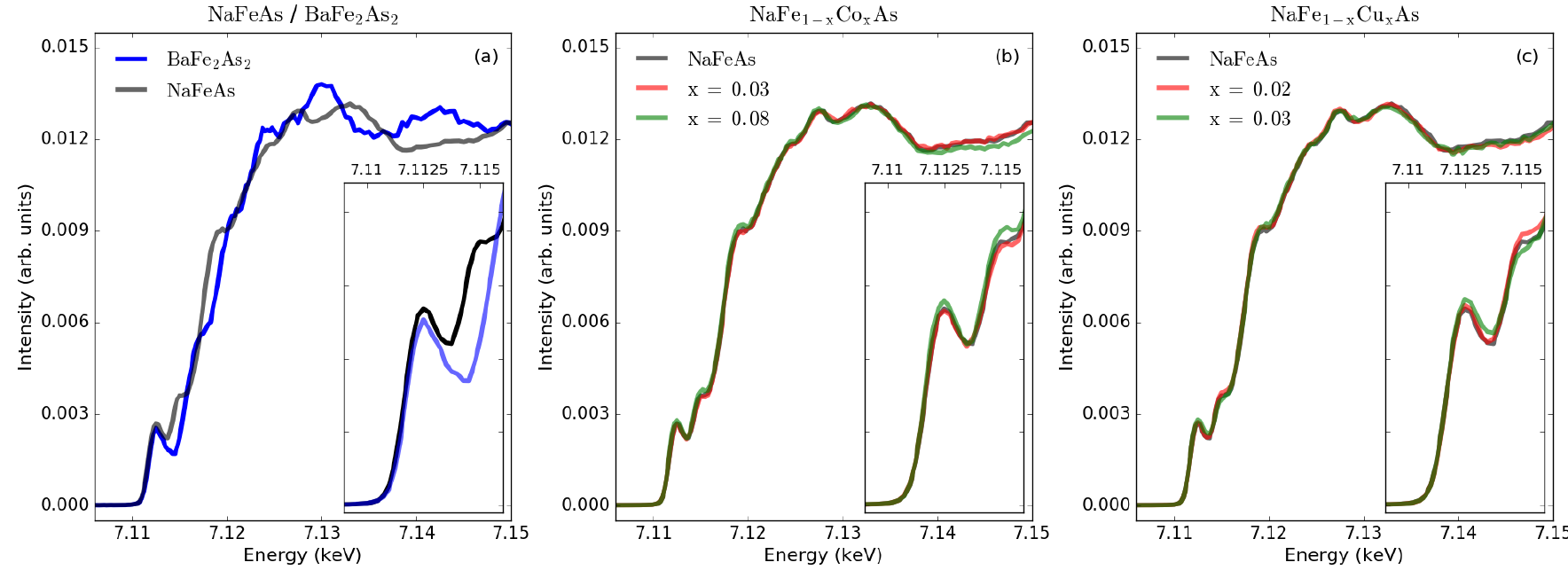}
\caption{\label{fig:fig2} (a) Fe-K edge XAS-PFY for $\mathrm{BaFe_2As_2}$ (blue solid line from Ref.~\cite{pelliciari_local_2017}) and $\mathrm{NaFeAs}$ (black solid line). Inset: Zoom on the pre-edge region for both compounds. (b) Fe-K edge XAS-PFY for $\mathrm{NaFe_{1-x}Co_{x}As}$ for x$=0$ (black solid line), 0.03 (red solid line), and 0.08 (green solid line). Inset: Zoom on the pre-edge region for all the compounds. (c) Fe-K edge XAS-PFY for $\mathrm{NaFe_{1-x}Cu_{x}As}$ for x$=0$ (black solid line), 0.02 (red solid line), and 0.03 (green solid line). Inset: Zoom on the pre-edge region for all the compounds.}
\end{figure*}

In $\mathrm{NaFe_{1-x}Co_xAs}$, high energy spin excitations have been observed to persist into the overdoped phase \cite{carr_electron_2016,pelliciari_intralayer_2016,zhang_anisotropic_2014,zhang_effect_2016,dai_magnetism_2012,dai_antiferromagnetic_2015}, with a decrease of their spectral weight, indicating that short-range magnetism permeates the phase diagram \cite{carr_electron_2016,pelliciari_intralayer_2016}.
Intriguing is the case of $\mathrm{NaFe_{1-x}Cu_{x}As}$ where a multitude of phenomena emerge: At low Cu doping (x $<0.3$) the typical superconducting dome appears in a very similar way to Co, but when high doping is performed a metal-to-insulator transition appears concomitantly to re-entrant AF \cite{tan_phase_2017,wang_phase_2013, song_mott_2016,liu_orbital-selective_2016,liu_localization_2015,cui_arpes_2013,matt_$mathrmnafe_0.56mathrmcu_0.44mathrmas$_2016}. This phenomenology has been linked to a Mott-selective phase transition at high Cu doping, connecting the physics of the Fe pnictides to the cuprates \cite{tan_phase_2017,wang_phase_2013, song_mott_2016,liu_orbital-selective_2016,liu_localization_2015,cui_arpes_2013,matt_$mathrmnafe_0.56mathrmcu_0.44mathrmas$_2016}.
These experimental pieces of evidence entail a different behavior of Co and Cu doping which is counterintuitive from an electron counting point of view. The electron counting works fairly well for the case of Co doping but is reversed for Cu doping where hole doping takes place \cite{song_mott_2016,liu_orbital-selective_2016,liu_localization_2015,tan_phase_2017,wang_phase_2013, song_mott_2016,liu_orbital-selective_2016,liu_localization_2015,cui_arpes_2013,matt_$mathrmnafe_0.56mathrmcu_0.44mathrmas$_2016}. From a magnetism perspective an open question is how the different doping of Co- and Cu- affects the local magnetic moment.

\begin{figure}
\includegraphics[scale=1]{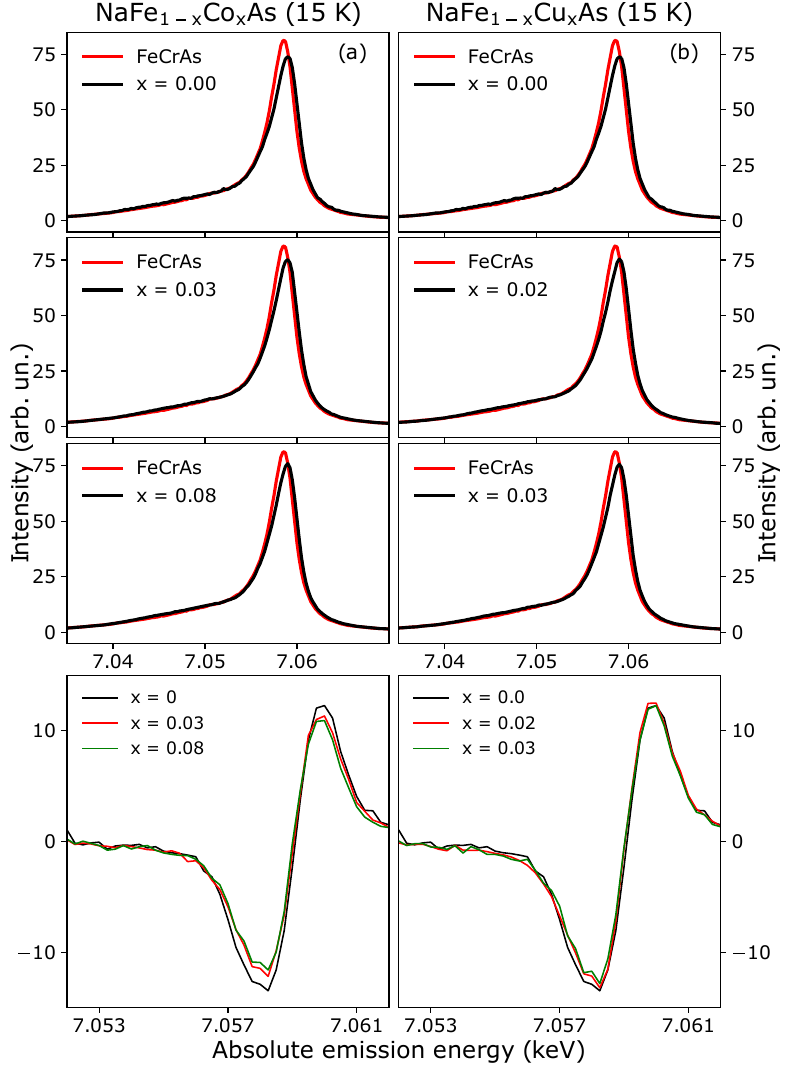}
\caption{\label{fig:fig3} (a) K$_{\beta}$ XES for $\mathrm{NaFe_{1-x}Co_{x}As}$ with x$=0$, 0.03, and 0.08 at 15~K and spectrum of FeCrAs used as a reference for calculating the difference. (b) K$_{\beta}$ XES for $\mathrm{NaFe_{1-x}Cu_{x}As}$ with x$=0$, 0.02, and 0.03 at 15~K  and spectrum of FeCrAs used for calculating the difference. The last row is indicating the relative difference spectra for $\mathrm{NaFe_{1-x}Co_{x}As}$ and $\mathrm{NaFe_{1-x}Cu_{x}As}$ using as reference FeCrAs.}
\end{figure}

Here, using of a combination of Fe-K edge x-ray absorption (XAS) and Fe-K$_\beta$ X-ray emission (XES), we report on the evolution of the local fluctuating magnetism of $\mathrm{NaFe_{1-x}Co_{x}As}$ (x $=0.03$ optimal doped T$_C=20$~K and x $=0.08$ overdoped  T$_C=6$~K) and $\mathrm{NaFe_{1-x}Cu_{x}As}$ (x $=0.02$ optimal doped  T$_C=12$~K and x $=0.03$ overdoped T$_C=5$~K). 
At 15~K, the XES experiments uncover a relative local fluctuating magnetic moment $\mu_{bare}$ in NaFeAs of 1.12 higher than $\mathrm{BaFe_2As_2}$ (normalized to 1.00). The frustration induced by a different distance between the As and Fe in NaFeAs \cite{zhang_effect_2016,yin_kinetic_2011,li_structural_2009} precludes the ordering of static magnetic moments in contrast to $\mathrm{BaFe_2As_2}$ leaving a high portion of spins fluctuating. Doping with Co slightly decreases $\mu_{bare}$ in optimal and overdoped samples, whereas in the case of Cu doping we observe very little modification of $\mu_{bare}$. Following electron counting arguments the optimal and overdoped samples of Co and Cu should have an equivalent number of carriers and consequently a similar spin state. Nonetheless, the different evolution of $\mu_{bare}$ for `isodoped' (meaning a nominal injection of the same number of carriers) Co- and Cu-doped samples reveals a different behavior. The former injects electrons into Fe affecting its spin state whereas the latter acts as a source of impurity. Lastly, we performed studies at 300~K and observed an increase of $\mu_{bare}$ in all the samples, which is indicative of the population of higher spin states at high temperature.

XAS at the K edges is an experimental technique sensitive to the oxidation state and local electronic symmetry of the atoms involved. In Fig.~\ref{fig:fig1}(a), we show the Fe-K edge XAS in partial fluorescence yield (PFY) for NaFeAs (black solid line) and $\mathrm{BaFe_2As_2}$ (blue solid line). The spectra display a very similar pre-edge peak at 7.1125 keV which can be ascribed to the FeAs hybridization peak observable thanks to the lack of inversion symmetry in the Fe tetrahedron leading to a projection of the 3d orbitals into the p ones \cite{lafuerza_evidence_2017}. At slightly higher energy we observe a second peak (7.11525 keV) ascribed to the sum of the dipole and quadrupole ($1s\rightarrow3d$) contribution \cite{bittar_co-substitution_2011,westre_multiplet_1997}. The main dipolar edge transition $1s\rightarrow4p$ starts at 7.116 keV. The oscillatory part of XAS at higher energy is different displaying that the two systems have a different local structure originating from different \textit{h}s (\textit{h}=1.358~\AA ~in $\mathrm{BaFe_2As_2}$ and \textit{h}=1.416~\AA~ in NaFeAs \cite{zhang_effect_2014,johnson_iron-based_2015,stewart_superconductivity_2011,dai_antiferromagnetic_2015}) of the As atoms. The longer bond length between Fe and As localizes the Fe electron in NaFeAs close to its site, enhancing its local magnetic moment.
In Fig.~\ref{fig:fig1}(b) and (c) we depict the XAS spectra of $\mathrm{NaFe_{1-x}Co_xAs}$ (x $=0.03$ and 0.08) and $\mathrm{NaFe_{1-x}Cu_xAs}$ (x $=0.02$ and 0.03) compared with the NaFeAs parent compound. The pre-edge peak at 7.1125 keV indicated in the zoom part of Fig.~\ref{fig:fig1}(b) and (c) changes very little with doping. The shape of the XAS spectra at higher energy is also pretty similar except for the shoulder appearing in the inset of Fig.~\ref{fig:fig1}(b) and (c) at 7.115 keV. This small variation may indicate a small difference in the structural environment due to a modification of the local structure happening with the substitution of transition metals of different sizes.

Previous Fe-K edge XAS experiments \cite{bittar_co-substitution_2011,baledent_stability_2012} on $\mathrm{BaFe_{2-x}Co_xAs_2}$ observed the independence of the XAS spectra upon doping. The invariance of XAS was interpreted as an indication that the doping is not affecting the valency of Fe atoms but that the injected electrons are rather localized around the dopant atoms which leads to impurity scattering of the itinerant electrons. The effect of impurity scattering has also been studied by theoretical calculations and confirmed by ARPES experiments indicating that the intralayer doped atoms localize the extra electrons close to them and induce a strong scattering, observed as a broadening of the bands \cite{wadati_where_2010,ye_extraordinary_2014}. Additionally, As-K edge XAS showed the effect of Co doping as well as hydrostatic pressure on the valence state of As above the optimally doped level. This indicates that the chemical pressure of Co doping affects the oxidation state of As leaving invariant the XAS at the Fe-K edge \cite{baledent_electronic_2015}. However this does not preclude that a small amount of electrons, too tiny to be detected by XAS, is transferred to the Fe ions leading to a modification of filling and consequent change of the spin state. 

XES is a photon-in photon-out spectroscopic technique that can be used to detect the spin state in materials \cite{peng_high-resolution_1994,bergmann_x-ray_2009, vanko_probing_2006, gretarsson_spin-state_2013, gretarsson_revealing_2011, ortenzi_structural_2015,simonelli_coexistence_2012,simonelli_temperature_2014,pelliciari_magnetic_2017,yamamoto_origin_2016,pelliciari_local_2017,vanko_spin-state_2013,vanko_temperature-_2006,sikora_1s2p_2012,rueff_pressure-induced_1999,rueff_magnetic_1999,rueff_short-range_2008}. The incoming photon ($h\nu$=7.140 keV) excites an Fe 1$s$ electron to the continuum with the creation of an unstable core-hole filled by the decay of an Fe 3$p$ electron and emission of a photon. This process is named K$_\beta$ emission. The final state of the system has a hole in the 3$p$ shell with a wavefunction partially overlapping with the 3$d$ states of the system which allows the sensitivity to the spin state \cite{vanko_probing_2006,tsutsumi_x-ray_1976,bergmann_x-ray_2009,vanko_spin-state_2013,glatzel_high_2005,peng_high-resolution_1994}. The K$_{\beta}$ emission line created with this mechanism is composed by a main peak, stemming from the sum of K$_{\beta_{1}}$ and K$_{\beta_{3}}$ and a satellite peak named K$_{\beta'}$ \cite{tsutsumi_x-ray_1976,vanko_spin-state_2013,vanko_probing_2006}. The K$_{\beta'}$ peak is directly sensitive to the spin state of the valence band, and using a proper calibration it is possible to extract the value of $\mu_\text{bare}$ \cite{bergmann_x-ray_2009, vanko_probing_2006, gretarsson_spin-state_2013, gretarsson_revealing_2011, ortenzi_structural_2015, simonelli_coexistence_2012, simonelli_temperature_2014,pelliciari_magnetic_2017,glatzel_high_2005,lafuerza_evidence_2017,peng_high-resolution_1994,peng_high-resolution_1994}. This spectroscopy probes the spin states with sensitivity to fast spin fluctuations overcoming the drawback of the quenching of the magnetic moment due to fast quantum fluctuations \cite{mannella_magnetic_2014,hansmann_dichotomy_2010,hansmann_uncertainty_2016}.

\begin{figure}
\includegraphics[scale=1]{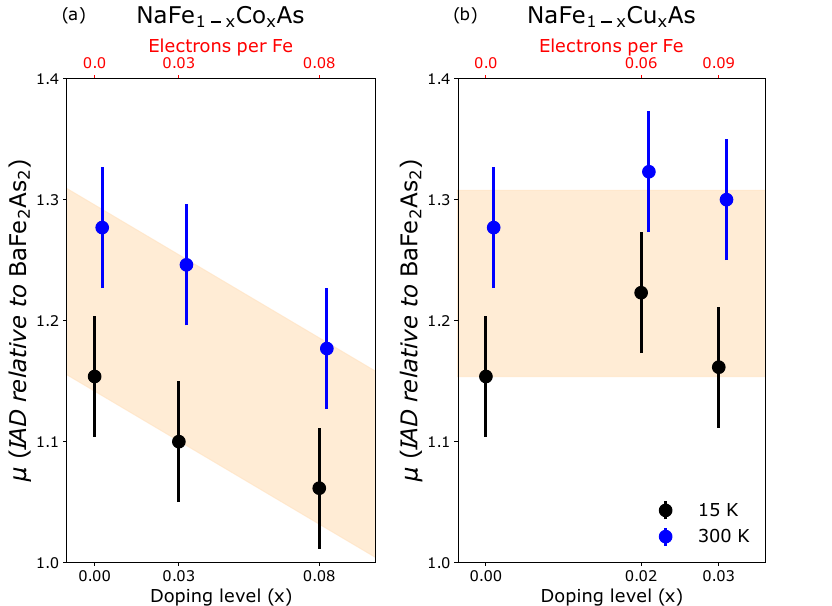} 
\caption{\label{fig:fig4} (a) Summary of the fluctuating local magnetic moment and IAD for $\mathrm{NaFe_{1-x}Co_{x}As}$ as a function of doping and temperature. (b) Summary of the fluctuating local magnetic moment and IAD for $\mathrm{NaFe_{1-x}Cu_{x}As}$ as a function of doping and temperature. The yellow shaded part is a guide for the eye.}
\end{figure}

In Fig.~\ref{fig:fig3}, we show the XES spectra collected for all the samples depicted as black lines and for the compound we have used for calibration, FeCrAs, which is represented by a solid red line. This compound has been used as a reference since it has no magnetic moment on the Fe sublattice and can hence be used as a calibrating sample \cite{wu_novel_2009,ishida_magnetic_1996,rau_hidden_2011,gretarsson_revealing_2011,gretarsson_resonant_2015}. In Fig.~\ref{fig:fig3}(a) and (b), we display the XES spectra for the $\mathrm{NaFe_{1-x}Co_xAs}$ and $\mathrm{NaFe_{1-x}Cu_xAs}$ samples at 15~K together with FeCrAs. The traces are composed of the K$_{\beta_{1}}$ and K$_{\beta_{3}}$ main line, and a satellite peak (K$_{\beta'}$) visible at lower energy as a shoulder. The shape of the spectra is similar for all the doping levels.
In the last row of Fig.~\ref{fig:fig3}(a) and (b), we show the difference spectra between the samples and the reference after normalization to the same area (see Supp. Mat. for details). The integration of the difference spectrum gives the integrated area difference (IAD) which is used to quantify the value of $\mu_{bare}$ shown in Fig.~\ref{fig:fig4} \cite{vanko_probing_2006,vanko_spin-state_2013,gretarsson_revealing_2011,gretarsson_spin-state_2013,pelliciari_local_2017,lafuerza_evidence_2017} (see Supp. Mat. for details).

We observe a slight decrease of the difference spectra upon Co doping as seen in Fig.~\ref{fig:fig2}(a), whereas the difference spectra for Cu doping are, within our error bars, unaffected by doping. This is an indication of a decrease of $\mu_{bare}$ in Co-doped samples and a constant spin state in Cu-doped specimens. The variation of $\mu_{bare}$ extracted using the Integrated Area Difference (IAD) is summarized in Fig.~\ref{fig:fig4}(a) and (b).
 From the values of IAD a clear change of spin state for the Co doping case can be inferred (Fig.~\ref{fig:fig4}(a)). This is not the case for the Cu-overdoped samples reported in Fig.~\ref{fig:fig4}(b) that shows very small difference spectra. This corroborates that the doping effect on the spin state of Co- and Cu-doped NaFeAs is different, with the former acting truly as electron doping, moving the formal filling of Fe from 3$d^6$ towards 3$d^7$ with a consequent decrease of the atomic spin. This variation is, however, not caught by the XAS measurements that show little change of the Fe valence state. It is important to say that for small variations of the Fe valence state the Fe-K edge XAS is not the most sensitive technique. 

For Cu doping instead, the spin state of Fe is unchanged, leading to the conclusion that Cu does not inject electrons into the system, in agreement with the idea that this  type of doping inserts impurity scattering centers without injecting electrons \cite{wadati_where_2010,ideta_dependence_2013,berlijn_transition-metal_2012}. For the specific case of Cu, it has been proposed that Cu plays the role of a hole donor \cite{liu_orbital-selective_2016,liu_localization_2015,song_mott_2016,matt_$mathrmnafe_0.56mathrmcu_0.44mathrmas$_2016} moving Fe from the formal 3$d^6$ configuration towards the half filling configuration 3$d^5$. This hole doping mechanism should consequently enhance the spin state as in $\mathrm{Ba_{1-x}K_xFe_2As_2}$ \cite{pelliciari_magnetic_2017,lafuerza_evidence_2017}. Such a doping effect is, however, not observed in our measurements, possibly implying that the regime of hole doping for Cu discussed by Ref.~\cite{song_mott_2016} appears only at very high Cu doping (x$>$0.3) and a different phenomenology has to be invoked at low doping.

An important consideration about the Cu-doped series concerns the origin of the ordering at high doping. NaFeAs has a low ordered moment ($\approx0.1\mu_B$ \cite{johnston_puzzle_2010,stewart_superconductivity_2011}), but a sizable fluctuating moment has been detected by INS, RIXS, and our present work \cite{zhang_effect_2014,pelliciari_intralayer_2016}. The appearance of an insulating phase \cite{wang_phase_2013,tan_phase_2017}, as well as a strong impurity potential and scattering, at high doping in $\mathrm{NaFe_{1-x}Cu_xAs}$ can, in principle, slow the fluctuating spins leading to the magnetic ordering observed in Ref.~\cite{song_mott_2016} and the enhancement of the magnitude of the ordered magnetic moment.

In LiFeAs, a photoemission study \cite{xing_anomaly_2014} revealed a similar dichotomy to our current report, where Co, Ni, and Cu played a different role as dopants for the parent compound. In this study it has been reported that Co and Ni substitution produces electron doping, injecting negative carriers. The effect of Cu doping is, however, different since the extra electrons localize close to the Cu atoms without being doped into the system \cite{xing_anomaly_2014}. 

A decrease of magnetic spectral weight in $\mathrm{NaFe_{1-x}Co_xAs}$ has been observed using INS and RIXS \cite{carr_electron_2016,pelliciari_intralayer_2016}. Noteworthy, INS and RIXS are momentum resolved techniques (which also probe different regions of the Brillouin zone) meanwhile XES is a local spectroscopy, so the information gathered from these spectroscopies is not necessarily the same, but rather projected in momentum or locally.
Similarly to our observation in $\mathrm{NaFe_{1-x}Co_xAs}$ the decrease of the fluctuating magnetism has been observed in $\mathrm{BaFe_{2-x}Co_xAs_2}$ by XES \cite{pelliciari_magnetic_2017,gretarsson_revealing_2011}, which is also in accord with the decrease of the magnetic spectral weight detected by INS \cite{wang_doping_2013}.

The IAD of NaFeAs at 300~K (Supp. Mat.) is increased compared to 15~K, indicating an increment of $\mu_{bare}$ of $\approx$7-10\% as summarized in Fig.~\ref{fig:fig4} (see Supp. Mat. for raw spectra). Interestingly, the enhancement of $\mu_{bare}$ at 300 K is observed also for Co and Cu doping ($\sim10-12$\%). This temperature evolution of the spin state has been reported in other Fe pnictides \cite{gretarsson_revealing_2011,gretarsson_spin-state_2013,pelliciari_magnetic_2017,pelliciari_local_2017} and represents a general feature of the Fe pnictides. It can be ascribed to the thermal population of high spin states and the interaction of the local spin with the electronic cloud that is affected by the temperature as described in the context of spin freezing \cite{werner_spin_2008,werner_satellites_2012,pelliciari_magnetic_2017,lafuerza_evidence_2017,georges_strong_2013,tytarenko_direct_2015}.

In conclusion, we performed Fe-K edge XAS and XES experiments on NaFeAs, unveiling the presence of a strong local fluctuating magnetic moment in NaFeAs which otherwise presents a low ordered magnetic moment. Experiments performed on $\mathrm{NaFe_{1-x}Co_xAs}$ [x $=0.03$ optimal doped (T$_C=20$~K) and x $=0.08$ overdoped (T$_C=6$~K)] and $\mathrm{NaFe_{1-x}Cu_xAs}$ [x $=0.02$ optimal doped (T$_C=12$~K) and x $=0.03$ overdoped (T$_C=5$~K)] uncovered a decrease of $\mu_{bare}$ in the case of Co doping and essentially constant $\mu_{bare}$ for Cu doping. This signals a different doping mechanism for these two transition metals and highlights the importance not only of the injection of carriers through doping but also the effect of impurity scattering.
We observed an increase of $\mu_{bare}$ in all the samples when raising the temperature to 300~K demonstrating that this phenomenology is a general feature of the Fe pnictides and should be accounted for in models to describe the magnetism of Fe pnictides.

\section*{Supplementary Materials}
See Supplementary Materials for additional information.

\section*{Availability of data}
The data that support the findings of this study are available from the corresponding authors upon reasonable request.

\begin{acknowledgments}
We thank Y. J. Kim and J. Mizuki for helpful discussions.
J.P. and T.S. acknowledge financial support through the Dysenos AG by Kabelwerke Brugg AG Holding, Fachhochschule Nordwestschweiz, and the Paul Scherrer Institut. J. P. acknowledges financial support by the Swiss National Science Foundation Early Postdoc. Mobility fellowship project number P2FRP2$\_$171824. The synchrotron radiation experiments have been performed at BL11XU of SPring-8 with the approval of the Japan Synchrotron Radiation Research Institute (JASRI) (Proposal No. 2016A3552). We thank D. Casa for fabrication of the Ge[620] analyzers installed at BL11XU of SPring-8. 
\end{acknowledgments}

\end{document}